# Optical Measurement of Mass Density of Biological Samples

*Conrad Möckel,[1,2,3] ‡ Jiarui Li,[4] ‡ Giulia Zanini,[4] Jochen Guck,[1,2,3] Giuliano Scarcelli[4]\**

[1]Max-Planck-Zentrum für Physik und Medizin, 91054 Erlangen, Germany

[2]Max Planck Institute for the Science of Light, 91058 Erlangen, Germany

[3]Friedrich-Alexander-Universität Erlangen-Nürnberg, 91058 Erlangen, German

[4]Fischell Department of Bioengineering, University of Maryland, 8278 Paint Branch Drive, College Park, MD 20742, USA

KEYWORDS: mass density, stimulated Brillouin scattering, biophysical sample, refractive index, biomechanics

**ABSTRACT:** Mass density is a vital property for improved biophysical understanding of and within biological samples. It is increasingly attracting active investigation, but still lacks reliable, non-contact techniques to accurately characterize it in biological systems. Contrary to popular belief, refractive index information alone is insufficient to determine a sample's mass density, as we demonstrate here theoretically and experimentally. Instead, we measured the nonlinear gain of stimulated Brillouin scattering to provide additional information for mass density estimation. This all-optical method reduces the estimation error tenfold, offering a more accurate and universal technique for mass density measurements.

**INTRODUCTION**

Mass density – defined as the total mass divided by the total volume in a sample – is increasingly recognized as an important property to quantify, especially in biophysical applications, as it is associated with cell functions and pathological conditions, and it is highly regulated during cell growth, differentiation, and metabolic activity[1]. Therefore, the development of measuring techniques for mass density is an area of active investigation[2]. While the measurement of volume is achievable with optical means in a straightforward manner, mass and mass density measurements require contact as they generally rely on buoyant force. The most widely used optical technique to estimate mass density at subcellular length scales is quantitative phase microscopy (QPM), an interferometric technique where density estimation relies on the experimental measurement of the refractive index, as will be discussed later. However, contrary to popular belief, the refractive index information is not sufficient to extract the mass density of a sample, as we will show here theoretically and experimentally; instead, additional information is needed for accurate retrieval of sample density.

Specifically, we show that such additional information can be provided by the measurement of the nonlinear gain of stimulated Brillouin scattering. With this all-optical method, the estimation error of mass density can be reduced tenfold for samples that cannot be processed by traditional methods, providing a more accurate and universal way for mass density measurements of biological samples.

**METHOD AND RESULTS**

Assuming a binary mixture of a solute (e.g., a protein, denoted by subscript s) and a solvent (e.g., water, denoted by subscript 0), the traditional estimation of the mass density from QPM in biological systems, e.g., in biological cells, relies on the assumption of the linearity of refractive index of the mixture $n$ with the solute concentration $c_s=(n-n_0)/\alpha$, governed by the Biot mixing rule of refractive indices[3], where $\alpha \equiv \partial n/\partial c_s$ is the customarily designated refractive index increment and $n_0$ is the refractive index of the solvent. By expressing the mass density of the mixture in terms of the solute and solvent concentrations $\rho=c_s+c_0$, one straightforwardly arrives at an expression for the mixture mass density $\rho$ in dependence on the (measured) mixture refractive index $n$ [2, 4]

$$\rho(n, \alpha, \theta) = \frac{n-n_0}{\alpha} + \rho_0 \cdot \left(1 - \theta \frac{n-n_0}{\alpha}\right), \qquad (1)$$

where $\rho_0$ is the solvent mass density and $\theta=V_{\text{dry}}/m_{\text{dry}}$ is the apparent specific volume (ASV) of the solute.

We apply Eq. (1) to a biologically relevant system, namely a cell. In the following, we consider a simplified, hypothetical cell that consists of cytoplasm, nucleoplasm, a nucleolus, lipid droplets, and protein aggregates. Based on this, we estimate the respective water concentrations of the different organelles using Eq. (1), the refractive index data of HeLa cells reported in [5] (cytoplasm, nucleoplasm and nucleolus) and [6] (specific protein aggregates) using the Biot equation. With these parameter values at our disposal, we employ Monte-Carlo simulations, as recently carried out in [4], to obtain the correlative distributions of refractive index and mass density of the different cell organelles. The results of the Monte-Carlo simulations are shown in Fig. 1, and the respective numerical values are given in Table S2. Cytoplasm, nucleoplasm, and nucleolus lie in a narrow range of mass densities (1.012–1.043 g/ml) and refractive indices (1.348–1.368), while protein aggregates and lipid droplets exhibit similarly high refractive index values (~1.463) but strikingly differ in density [$\rho_{\text{PA}}=(1.188\pm0.023)$ g/ml, $\rho_{LD}$=0.90 g/ml]. This implies that biological matter within the above refractive index range has a roughly estimated density of ~1.028 g/ml, ignoring the details about macromolecular composition. Otherwise, one would need additional information about their composition to avoid under/overestimating the mass density significantly. (See details in Note S1 of the Supporting Information.)

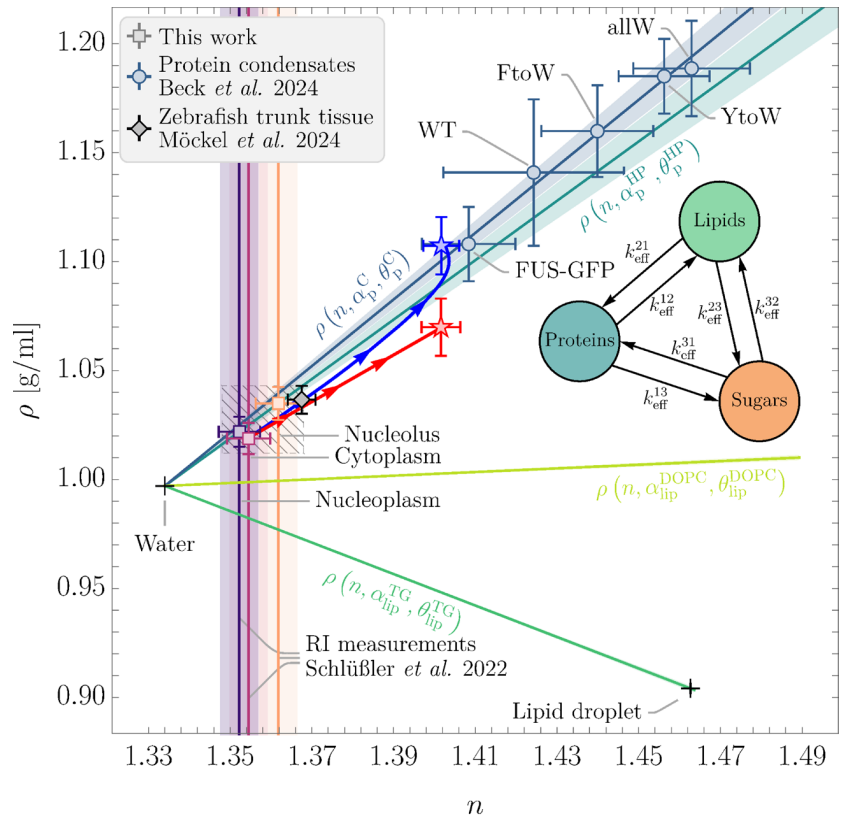

**Figure 1.** Mass density $\rho$ in dependence of the refractive index $n$ for different cell organelles, protein aggregates and zebrafish trunk tissue. The colorful squares indicate the simulation results of the different cell organelles described in the main text (Median ± 95% CI), the dark blue circles show values (Median ± 95% CI) of protein condensates of different proteins based on the experimental refractive index data of[6], and the black diamond indicates the values for larval zebrafish tissue from[4] (Median ± 95% CI). The vertical colorful bands indicate refractive index measurements of the respective cell organelles of [5] for the respective organelles. The full lines

and corresponding bands indicate Eq. (1) for the cases of condensate proteins (C), the human proteome (HP), DOPC and triglyceride (TG), respectively and the full lines with arrows indicate the two hypothetical hyper osmotic shock scenarios described in the main text; A) (blue) and B) (red), using the minimal three-component reaction network displayed in the inset to the right. The direction of the arrows indicates a decrease in water concentration and the final time point is indicated by a star (Median ± 95% CI).

So far, we have considered cases where the composition of the biological matter is maintained over time. While this is certainly a valid assumption for cells and tissues that remain homeostatic[7], in the next step, we will investigate cases where the macromolecular composition changes over time (e.g., due to actively driven macromolecular conversions; see[8] for a review). We hypothesize that such scenarios might be relevant for studies of cases where the cell/tissue homeostasis is disturbed, causing stress responses and potentially leading to e.g. an inflammatory response (for a review, see [9]). To outline the implications of such scenarios on the mass density-refractive index relationship, despite a lack of quantitative empirical data, we describe this problem using a minimal reaction network including three types of macromolecules, abundant in living matter, namely proteins, lipids and sugars[8,10], shown in Fig. 1. We assume that each type of macromolecule can be effectively converted into another with the respective conversion rates $k^{ij}_{eff}$, where the $i \neq j = \{1,2,3\}$ denote the individual macromolecules. By setting the initial conditions concordant to the composition of cytoplasm, described earlier, we numerically solve the resulting set of coupled rate equations for the two cases; A) $k^{ij}_{eff} \neq 0$, i.e., macromolecular conversions, as described above and B) $k^{ij}_{eff} = 0$, i.e., no macromolecular conversions, using the NDSolve function in Mathematica [11], as shown in Fig. 1 and Fig. S2. As expected, the resulting correlative mass density-refractive index relationship is non-linear for case A and linear for case

B. While the difference between the two cases is not significant for the refractive index range stated above (1.348–1.368), i.e., at high water concentrations, the effect of the (slow) macromolecular conversion becomes more apparent for lower water concentrations. In fact, the difference of the final refractive indexes is marginal, but the difference in the mass densities is immanent. This, in turn, strongly motivates the necessity to measure the mass density directly. Going forward, correlative mass density and refractive index measurements could lead to quantification of effective macromolecular conversion rates. (See details in Note S2 of the Supporting Information.)

An exemplary model system that obeys the relationship of Eq. (1) is a solution of bovine serum albumin (BSA) in phosphate-buffered saline. As shown in Fig. 2(a), density (measured as mass/volume with the use of an analytical balance) and refractive index (measured with an Abbe refractometer) exhibits a linear relationship which can be well fitted by Eq. (1), giving a BSA dry density $\rho_{dry}$=1.32 g/ml, consistent with the average value for proteins.[12-14] The corresponding percent error of this linear fitting to the measured protein concentration, illustrated in Fig. 2(b), is always under 0.5% demonstrating that in these scenarios the measurement of refractive index is sufficient to extract mass density. However, scenarios where Eq. (1) does not describe the physics of the system are extremely common[15] and span from biological systems whenever lipid and protein relative concentrations are not constant even to simpler systems, such as water-alcohol mixtures. To illustrate this scenario, we chose a simple model system of water-methanol mixtures with methanol molar fractions ranging from 0 to 1. Fig. 2(c) shows the directly measured density $\rho_{meas}$, and the linear estimated values $\rho_{lin}$, at different methanol molar fractions, represented as solid stars and the dashed line, respectively. Assuming the linearity of density with refractive index as $\rho_{lin}=\rho_w+\beta(n-n_w)$, with $\beta=(\rho_m-\rho_w)/(n_m-n_w)$ and where subscript w stands for water and m for methanol, leads to up to 50% density overestimation, as shown in Fig. 2(d). This phenomenon is well understood in the literature[16]: the mixing behavior deviates from the linear case governed by volume additivity because the composition affects the thermodynamic properties and induces a (negative) excess molar volume. It is obvious that in these scenarios, measuring only refractive index would lead to dramatic errors in the estimation of sample density.

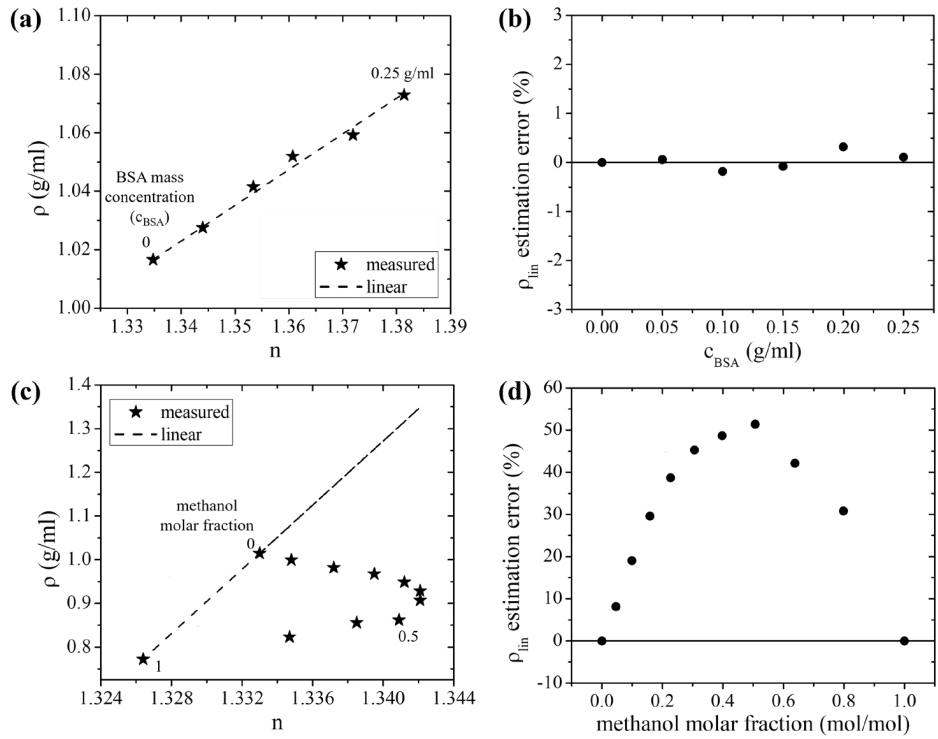

**Figure 2.** (a) Measured density (solid stars) of BSA solutions with BSA concentrations ranging from 0 to 0.25 g/ml (m/v), was chosen to mimic the cellular environment, as a function of measured refractive index. The dashed line represents the linear fit of the data using Eq. (1) and the tabulated value for protein dry mass density $\rho_{dry}$=1.32 g/ml. (b) Percent error in estimating the mass density of BSA solutions from linear approximation (solid circles). (c) Measured density (solid stars) of water-methanol solutions (at indicated molar fractions) as a function of measured refractive index. The dashed line represents the mass density derived assuming the linearity of density with refractive index as $\rho_{\text{lin}}=\rho_w+\beta(n-n_w)$. (d) Percent error in estimating the mass density of water-methanol mixtures from linear approximation (solid circles).

It is thus important to investigate if there are additional optical measurements that can augment the refractive index information for an accurate and universal retrieval of mass density in all types of samples. Here, we show that the stimulated Brillouin scattering (SBS) can provide such information. Indeed, Brillouin scattering arises from the interaction of acoustic phonons, whose propagation is governed by density fluctuations, and optical photons, whose propagation is governed by the refractive index. Namely, the strength of Brillouin scattering interaction is basically governed by how efficiently a density modulation translates into a diffraction grating, and thus provides an additional independent relation of density and refractive index. According to the Clausius–Mossotti relation, the relationship between the density $\rho$ and the dielectric constant $\varepsilon$, which for non-magnetic materials is equal to $n^2$, can be expressed as

$$\varepsilon = \left(1 + \frac{2\rho\alpha}{3\varepsilon_0}\right)\left(1 - \frac{\rho\alpha}{3\varepsilon_0}\right)^{-1}, \qquad (2)$$

where $\alpha$ is the molecular polarizability and $\varepsilon_0$ is the permittivity of free space. The electrostrictive constant $\gamma_e$ can be simplified as

$$\gamma_e = \rho\frac{\partial \varepsilon}{\partial \rho} = \frac{\rho\alpha}{\varepsilon_0}\left(1 - \frac{\rho\alpha}{3\varepsilon_0}\right)^{-2} = \frac{1}{3}(\varepsilon - 1)(\varepsilon + 2) \approx \frac{1}{3}(n^2 - 1)(n^2 + 2). \qquad (3)$$

which is only related to the refractive index. In this case, the mass density, which is contained in the Brillouin gain of the SBS spectrum, can be independently derived from the spectral information and the refractive index[17]:

$$\rho_{SBS} = \frac{4\pi\gamma_e^2}{c\lambda^3 v_B \Gamma_B g_B} = \frac{4\pi}{9c\lambda^3} \frac{(n^2-1)^2(n^2+2)^2}{v_B \Gamma_B g_B}, \qquad (4)$$

where $c$ is the speed of light, $\lambda$ is the wavelength of the incident light, $v_B$ is the characteristic Brillouin shift of the scattered light (on the order of GHz), and $\Gamma_B$ and $g_B$ represent the linewidth and line-center gain factor, respectively. The detailed derivation is shown in Note S3 of the Supporting Information. To experimentally demonstrate the validity of our insight, we built an SBS setup similar to previous works. [18, 19] [Fig. S3(a)]. Briefly, two counter-propagating laser beams, a pump and a probe, centered at 780 nm, are focused and overlapped at the sample. The scattered light is sensitively detected as gain in the transmitted probe using a lock-in detection scheme, and spectrally analyzed by scanning the frequency difference between the two lasers, $\Delta v$, around the characteristic Brillouin frequency [Fig. S3(b)].

Then we went back to analyzing the two systems in Fig. 2 using the additional information provided by SBS. Specifically, we estimated the density using Eq. (4) and the previously measured values of refractive index. The densities retrieved from the SBS spectra, $\rho_{SBS}$, as shown in Figs. 3(a) and 3(c), are in good agreement with the measured values, $\rho_{meas}$, as shown in Figs. 2(a) and 2(c), as expected: in BSA solutions, the density linearly increases with the BSA concentration, $c_{BSA}$, and agrees with the $\rho_{lin}$ value calculated using Eq. (1) and $\rho_{dry}$=1.32 g/ml. The corresponding error between the SBS estimated densities, $\rho_{SBS}$, and the directly measured values, $\rho_{meas}$, is shown in Fig. 3(b), always within 2% of the true value. The clear demonstration of the superiority of this method is in the water-methanol solutions. The densities retrieved by SBS reproduce the proper relationship to refractive index, Fig. 3(c) and the absolute estimation error of $\rho_{SBS}$ from SBS estimation is never more than 5%, Fig. 3(d). Importantly, the uncertainties displayed by these measurements are all experimental due to the potential error in quantifying overlap and residual beam absorption, which can be further adjusted in the future.

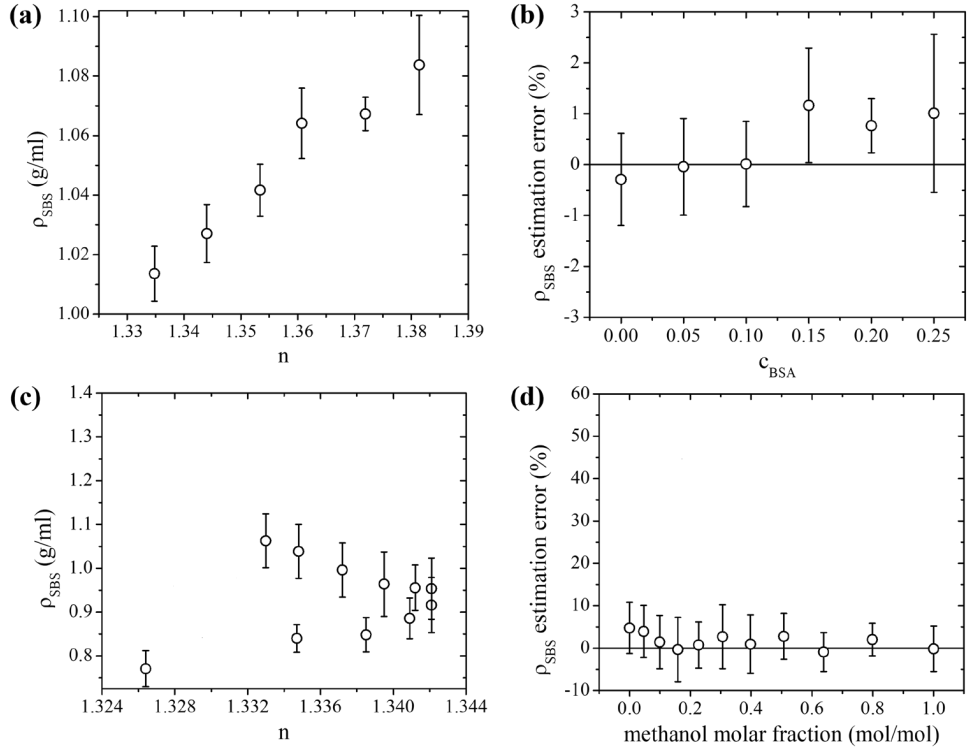

**Figure 3.** (a) and (c) Mass density estimation from SBS measurements [Eq. (4) using the measured value of refractive index] for BSA solutions and water-methanol mixtures, respectively. (b) and (d) Percent error in estimating the mass density of BSA solutions and water-methanol mixtures, respectively, through SBS measurements. For water-methanol mixtures, 100 spectra were analyzed, and their fit parameters averaged. For BSA mixtures, instead, multiple replicas of 100-repetition experiments were used to estimate more precisely the small density variation across the different BSA concentrations, which is four times smaller than in the water-methanol case. Spectra were scanned in 50 ms, and powers at the sample were kept around 10 mW for the probe beam, and around 100-150 mW for the pump beam, lowering it below 100 mW in samples with high methanol content to prevent evaporation.

The accurate estimation of refractive index and density is also important for the nascent biomechanics field powered by Brillouin microscopy, where the quantification of the longitudinal elastic modulus $M'$ is related to the Brillouin scattering shift $v_B$ via

$$M' = \frac{\rho}{n^2}\frac{\lambda^2}{4}v_B^2, \quad (5)$$

which includes the ratio between density, $\rho$, and refractive index, $n$. Previous studies have shown exactly the same behavior of Figs. 2 and 3. I.e., for certain samples, where the assumption of the monotonically increasing relationship between density and refractive index described in Eq. (1) is valid, the longitudinal modulus estimation can ignore the factor $\rho/n^2$, leading to very small errors (< 1%)[20, 21]. However, whenever the two-substance mixture model and Eq. (1) do not hold, as in lipid droplets within cells, the lack of index/density information leads to >25% inaccuracy in the estimation of the longitudinal modulus[5]. We therefore asked if the refractive index information and the measurement of SBS gain could resolve this issue too, using the same samples of Figs. 2 and 3, but focusing on the quantification of longitudinal modulus $M'$. For BSA mixtures, given the good linearity assumption of density and index, the $\rho/n^2$ factor is constant across the different concentrations, with an average value of 0.57 g/ml [Fig. 4(a)], consistent with previous assumptions in biological environments [20, 21]. This leads to modulus estimation error results within 2% regardless of whether density/index are measured, calculated, or assumed constant [Fig. 4(b)]. The case of water-methanol mixtures is much more interesting: the factor $\rho/n^2$ changes by up to 30% across the different compositions [Fig. 4(c)] which leads to the necessity of quantifying both refractive index and density. As shown in Fig. 4(d), assuming an average value of 0.5 g/ml as a constant in the modulus estimation, an error of $M'$ estimation up to 15% arises (triangles). The situation is worsened if only the measurement of refractive index is performed and linearity with density is assumed, in which case this error increases up to 50% (solid dots). Adding the measurement of the density from the SBS method largely resolves the issue, leading to errors lower than 5%.

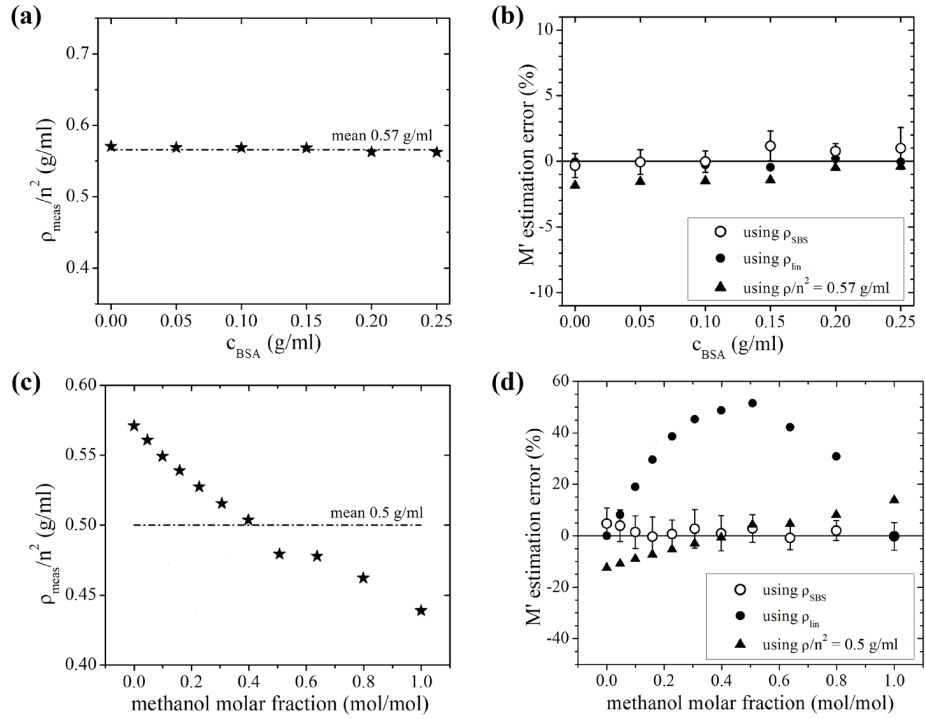

**Figure 4.** (a) and (c) Ratio $\rho/n^2$ for BSA solutions and water-methanol mixtures, respectively, calculated using measured values for both $\rho$ and $n$ (solid stars). The dash-dotted line represents the mean value over the range of analyzed BSA concentrations and water-methanol molar fractions. (b) and (d) Percent error in estimating the longitudinal modulus $M'$ through Eq. (5) using the density value derived from SBS (empty circles), the one derived from the linear approximation (solid circles), or using a constant $\rho/n^2$ ratio (solid triangles).

**CONCLUSION**

In summary, we have demonstrated that the all-optical estimation of mass density is not universally possible by simply mapping the refractive index with interferometric methods, because often in biophysical systems, the assumption of linearity between refractive index and density does not hold. Importantly, combining co-localized mapping of refractive index, obtained via interferometry, and additional techniques that probe a different dependence on refractive index and density, as demonstrated here by SBS, all-optical estimation of mass density can be achieved in biological samples with high accuracy.

ASSOCIATED CONTENT

**Supporting Information**

The Supporting Information is attached to this document.

Spatial distribution of relevant quantities of the cellular organelles considered in this study and results of the Monte-Carlo simulations (Figure S1); simulated concentration results using the minimal reaction network introduced in the main text for the hypothetical case of a hyperosmotic shock (Figure S2); schematic of the SBS setup and characteristic SBS spectrum of water (Figure S3); Refractive index increments $\alpha$ and apparent specific volumes $\theta$ for the different molecules employed in this study (Table S1); list of parameters for different cell organelles employed in this study (Table S2); minimal reaction network parameters for two different hypothetical scenarios (Table S3); parameter values and cell model when the composition of biological matter is maintained over time (Note S1); parameter values and cell model when macromolecular composition changes over time (Note S2); and derivation of the density estimated from the SBS spectrum (Note S3) (PDF)

AUTHOR INFORMATION

**Corresponding Author**

Giuliano Scarcelli − Fischell Department of Bioengineering, University of Maryland, 8278 Paint Branch Drive, College Park, MD 20742, USA; Email: scarc@umd.edu

**Author Contributions**

‡These authors contributed equally.

**Funding Sources**

This research has been funded in part by the National Institutes of Health (R01GM157084, R21CA258008, R01EY032537). JG acknowledges core funding by the Max Planck Society and the HORIZON-EIC Pathfinder Project DISRUPT (#101099663).

**Notes**

The authors declare no competing financial interest.

ABBREVIATIONS

QPM, quantitative phase microscopy; ASV, apparent specific volume; BSA, bovine serum albumin; SBS, stimulated Brillouin scattering.

REFERENCES

(1) Neurohr, G. E.; Amon, A., Relevance and regulation of cell density. *Trends Cell Biol* **2020,** *30* (3), 213-225.

(2) Zangle, T. A.; Teitell, M. A., Live-cell mass profiling: an emerging approach in quantitative biophysics. *Nat Methods* **2014,** *11* (12), 1221-1228.

(3) Heller, W., Remarks on refractive index mixture rules. *J. Phys. Chem.* **1965,** *69* (4), 1123-1129.

(4) Möckel, C.; Beck, T.; Kaliman, S.; Abuhattum, S.; Kim, K.; Kolb, J.; Wehner, D.; Zaburdaev, V.; Guck, J., Estimation of the mass density of biological matter from refractive index measurements. *Biophys. Rep.* **2024,** *4* (2), 100156.

(5) Schlüßler, R.; Kim, K.; Nötzel, M.; Taubenberger, A.; Abuhattum, S.; Beck, T.; Müller, P.; Maharana, S.; Cojoc, G.; Girardo, S., Correlative all-optical quantification of mass density and mechanics of subcellular compartments with fluorescence specificity. *Elife* **2022,** *11*, e68490.

(6) Beck, T.; van der Linden, L.-M.; Borcherds, W. M.; Kim, K.; Schlüßler, R.; Müller, P.; Franzmann, T.; Möckel, C.; Goswami, R.; Leaver, M.; Mittag, T.; Alberti, S.; Guck, J., Optical characterization of molecular interaction strength in protein condensates. *Mol. Biol. Cell.* **2024,** *35* (12), ar154.

(7) Liu, X.; Oh, S.; Kirschner, M. W., The uniformity and stability of cellular mass density in mammalian cell culture. *Front. Cell Dev. Biol* **2022,** *10*, 1017499.

(8) Judge, A.; Dodd, Michael S., Metabolism. *Essays Biochem.* **2020,** *64* (4), 607-647.

(9) Chovatiya, R.; Medzhitov, R., Stress, Inflammation, and Defense of Homeostasis. *Mol. Cell* **2014,** *54* (2), 281-288.

(10) Rollin, R.; Joanny, J.-F.; Sens, P., Physical basis of the cell size scaling laws. *Elife* **2023,** *12*, e82490.

(11) Inc., W. R., *Mathematica Version 12.2*. Wolfram Research. Inc.: Champaign, 2020.

(12) Schlüßler, R.; Möllmert, S.; Abuhattum, S.; Cojoc, G.; Müller, P.; Kim, K.; Möckel, C.; Zimmermann, C.; Czarske, J.; Guck, J., Mechanical mapping of spinal cord growth and repair in living zebrafish larvae by Brillouin imaging. *Biophys. J.* **2018,** *115* (5), 911-923.

(13) Harpaz, Y.; Gerstein, M.; Chothia, C., Volume changes on protein folding. *Structure* **1994,** *2* (7), 641-649.

(14) Zhao, H.; Brown, P. H.; Schuck, P., On the distribution of protein refractive index increments. *Biophys. J.* **2011,** *100* (9), 2309-2317.

(15) Barr, E. S., Concerning index of refraction and density. *Am. J. Phys* **1955,** *23* (9), 623-624.

(16) Herráez, J. V.; Belda, R., Refractive indices, densities and excess molar volumes of monoalcohols+ water. *J Solution Chem* **2006,** *35*, 1315-1328.

(17) Boyd, R. W., *Nonlinear optics*. Springer International Publishing: Cham, 2008.

(18) Remer, I.; Shaashoua, R.; Shemesh, N.; Ben-Zvi, A.; Bilenca, A., High-sensitivity and high-specificity biomechanical imaging by stimulated Brillouin scattering microscopy. *Nat Methods* **2020,** *17* (9), 913-916.

(19) Zanini, G.; Scarcelli, G., Localization-assisted stimulated Brillouin scattering spectroscopy. *APL photonics* **2022,** *7* (5).

(20) Scarcelli, G.; Pineda, R.; Yun, S. H., Brillouin optical microscopy for corneal biomechanics. *J. Invest. Dermatol.* **2012,** *53* (1), 185-190.

(21) Scarcelli, G.; Polacheck, W. J.; Nia, H. T.; Patel, K.; Grodzinsky, A. J.; Kamm, R. D.; Yun, S. H., Noncontact three-dimensional mapping of intracellular hydromechanical properties by Brillouin microscopy. *Nat Methods* **2015,** *12* (12), 1132-1134.

# Supporting Information for

# Optical Measurement of Mass Density of Biological Samples

*Conrad Möckel,[1,2,3] ‡ Jiarui Li,[4]‡ Giulia Zanini,[4] Jochen Guck,[1,2,3] Giuliano Scarcelli[4]\**

[1]Max-Planck-Zentrum für Physik und Medizin, 91054 Erlangen, Germany

[2]Max Planck Institute for the Science of Light, 91058 Erlangen, Germany

[3]Friedrich-Alexander-Universität Erlangen-Nürnberg, 91058 Erlangen, German

[4]Fischell Department of Bioengineering, University of Maryland, 8278 Paint Branch Drive, College Park, MD 20742, USA

‡These authors contributed equally.

\* Corresponding author Email: scarc@umd.edu

# Contents

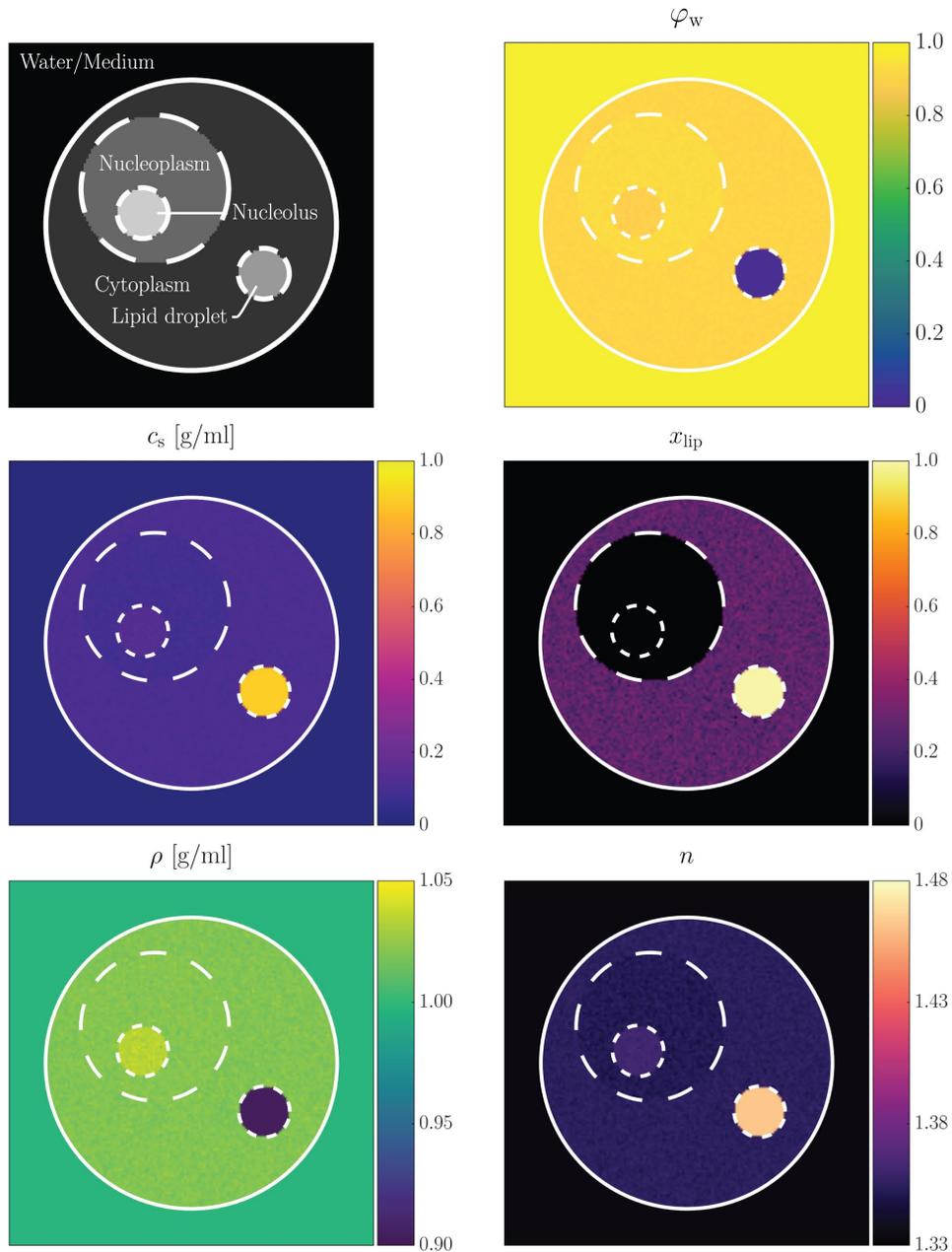

**Figure S1.** Spatial distribution of relevant quantities (water volume fraction $\varphi_w$, solute concentration $c_s$, lipid-to-protein volume ratio $x_{lip}$, mass density $\rho$ and refractive index $n$) of the cellular organelles considered in this study and results of the Monte-Carlo simulations, following[1].

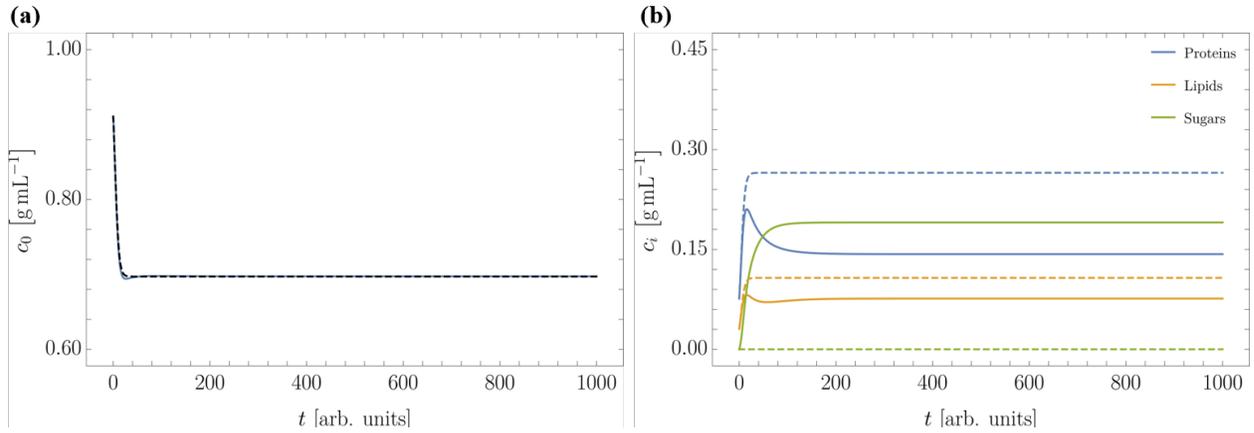

**Figure S2.** Simulated concentration results using the minimal reaction network introduced in the main text for the hypothetical case of a hyperosmotic shock. (a) Water concentration $c_0$ in dependence on simulation time for the case of no macromolecular conversions (black, dashed) and macromolecular conversions (blue, solid). (b) Solute concentrations $c_i$ in dependence on simulation time, for the case of no macromolecular conversions (dashed) and macromolecular conversions (solid).

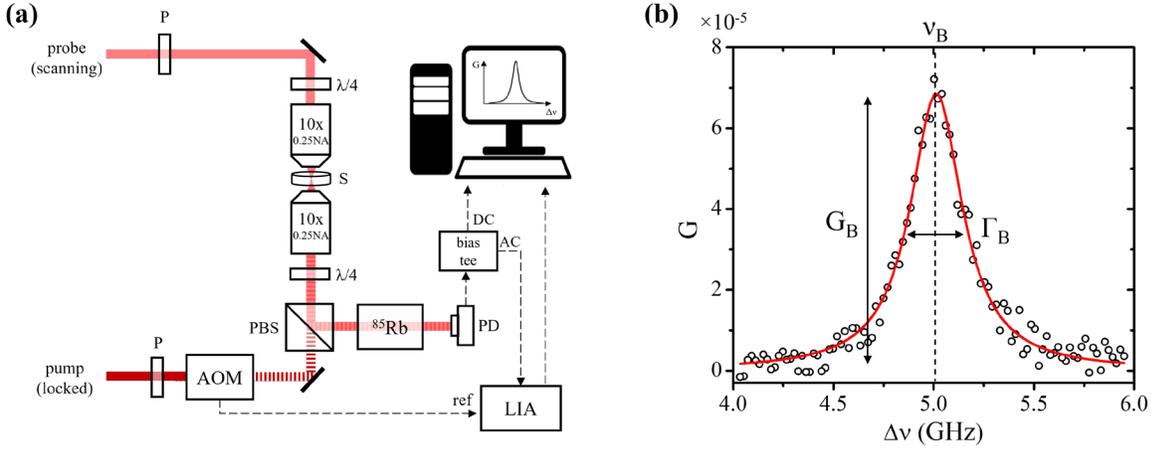

**Figure S3.** (a) Schematic of the SBS setup. P: polarizer, AOM: acousto-optic modulator, PBS: polarizing beam splitter, $\lambda/4$: quarter-wave plate, NA: numerical aperture, S: sample, $^{85}$Rb: rubidium vapor cell used as notch filter, PD: photodetector, LIA: lock-in amplifier. (b) Characteristic SBS spectrum of water, fitted with a Lorentzian function (red line) to extract Brillouin shift $\nu_B$, linewidth $\Gamma_B$, and gain $G_B$.

**Table S1.** Refractive Index Increments α and Apparent Specific Volumes θ for the Different Molecules Employed in this Study.

| (Macro) molecules | $\alpha$(ml/g) | $\theta$(ml/g) |
|---|---|---|
| **Proteins**[a] | | |
| Human Proteome | 0.197±0.004 | 0.734±0.012 |
| Protein condensates | 0.208±0.007 | 0.685±0.011 |
| FUS | 0.200 | 0.704 |
| WT | 0.203 | 0.681 |
| All W | 0.216 | 0.680 |
| F2W | 0.212 | 0.676 |
| Y2W | 0.207 | 0.683 |
| **Lipids** | | |
| Phospho-lipids | | |
| DOPC[b] | 0.142 | 0.925 |
| Neutral lipids | | |
| Triaglyceride[c] | 0.154 | 1.107 |
| **Sugars** | | |
| D-Glycogen[d] | 0.150±0.009 | 0.63±0.02 |

[a] Values computed based on the respective amino acid sequence as described in [1,2]

[b] Refractive index increment $\alpha$ computed based on the molar refractivity from [3] and the ASV $\theta$ of [4], see [1] for further information

[c] Refractive index increment $\alpha$ computed based on the molar refractivity from [5] and the ASV $\theta$ of [6], see [1] for further information

[d] Refractive index increment $\alpha$ computed based on the molar refractivity from [7] and the ASV $\theta$ of [8], see [1] for further information

**Table S2.** List of Parameters for Different Cell Organelles Employed in this Study.*

|  | Cytoplasm | Nucleoplasm | Nucleolus | Lipid droplet |
|---|---|---|---|---|
| $\rho$(g/ml) | 1.019±0.007 | 1.022±0.007 | 1.035±0.008 | 0.904 |
| $n$ | 1.355±0.004 | 1.352±0.004 | 1.362±0.004 | 1.463 |
| $c_0$(g/ml) | 0.908±0.011 | 0.929±0.009 | 0.894±0.010 | 0.007 |
| $c_1^{HP}$(g/ml) | 0.080±0.005 | 0.092±0.013 | 0.141±0.014 | 0 |
| $c_2^{DOPC}$(g/ml) | 0.031±0.006 | 0 | 0 | 0 |
| $c_2^{TG}$(g/ml) | 0 | 0 | 0 | 0.904 |
| $c_3$ (g/ml) | 0 | 0 | 0 | 0 |

* $\rho$ mass density, $n$ refractive index, $c_0$ water concentration, $c_1$ protein concentration, $c_2$ lipid concentration and $c_3$ sugar concentration.

**Table S3.** Minimal Reaction Network Parameters for Two Different Hypothetical Scenarios*

|    | $m_0^0$ | $m_1^0$ | $m_2^0$ | $m_3^0$ | $k_0$ | $m_0^\infty$ | $k_{eff}^{12}$ | $k_{eff}^{21}$ | $k_{eff}^{23}$ | $k_{eff}^{32}$ | $k_{eff}^{31}$ | $k_{eff}^{13}$ |
|----|---------|---------|---------|---------|-------|--------------|----------------|----------------|----------------|----------------|----------------|----------------|
| a) | 2 | 0.167 | 0.0677 | 0 | 0.25 | 0.11 | 0 | 0 | 0.0025 | 0.010 | 0.020 | 0.015 |
| b) | 2 | 0.167 | 0.0677 | 0 | 0.25 | 0.11 | 0 | 0 | 0 | 0 | 0 | 0 |

*$m_0^0$ initial water mass, $m_1^0$ initial protein mass, $m_2^0$ initial lipid mass, $m_3^0$ initial sugar mass, $k_0$ water rate, $m_0^\infty$ final water mass, $k_{eff}^{12}$ effective protein to sugar conversion rate, $k_{eff}^{21}$ effective sugar to protein conversion rate, $k_{eff}^{23}$ effective sugar to lipid conversion rate, $k_{eff}^{32}$ effective lipid to sugar conversion rate, $k_{eff}^{31}$ effective sugar to protein conversion rate, and $k_{eff}^{13}$ effective protein to sugar conversion rate.

**Note S1. Parameter values and cell model when the composition of biological matter is maintained over time:**

Biological cells exhibit tremendously complex macromolecular composition with varying concentrations across different sub-cellular organelles [9], rendering a detailed a microscopic analysis a challenging endeavor. To simplify this complexity, we consider a hypothetical cell that consists of cytoplasm, nucleoplasm, a nucleolus, lipid droplets, and protein aggregates, as shown in Fig. S1. Based on [10] and references therein, we assume that the refractive index increment and apparent specific volume (ASV) of the proteins in the cytoplasm and solutes present in the nucleoplasm and nucleolus are identical to the average values of the human proteome, for which the values may be computed based on the amino acid sequences of the individual proteins [1,2]. Following this approach, [11] computed refractive index increments and ASVs for the specific phase separating proteins under study therein, which we adopt here (see Table S1). Taking into account the stimulated Raman spectroscopy data of HeLa cells reported in [12], we assume that the cytoplasm additionally contains the membrane phospholipid DOPC and estimate the protein-to-lipid mass ratio as $y_p = c_p/(c_p + c_{lip}^{DOPC}) = 0.72 \pm 0.08$ (Median $\pm$ 95% CI). Additionally, we assume that the lipid droplets consist only of triglyceride. Based on this, we estimate the respective water concentrations of the different organelles using Eq. (1), the refractive index data of HeLa cells reported in [13] (cytoplasm, nucleoplasm and nucleolus) and [11] (specific protein aggregates) using the Biot equation. We find that, consistent with previous findings [14,15], the nucleus has a higher water concentration than the cytoplasm, nucleolus and protein aggregates ($c_0^{PA} < c_0^{nucleo} < c_0^{cyt} < c_0^{nuc}$, where subscript PA stands for protein aggregates), as shown in Table S2.

With these parameter values at our disposal, we employ Monte-Carlo simulations, as recently carried out in [1], to obtain the correlative distributions of refractive index and mass density of the different cell organelles. The results of the Monte-Carlo simulations are shown in Fig. 1 and the respective numerical values are given in Table S2.

By this analysis we find that the estimated mass density of the nucleolus [$\rho_{\text{nucleo}}$=(1.035±0.008) g/ml] is significantly higher than the mass densities of the cytoplasm and nucleoplasm [$\rho_{\text{cyt}}$ =(1.019±0.007) g/ml, $\rho_{\text{nuc}}$=(1.022±0.007) g/ml], respectively, which is due to the lower water concentrations present in the cytoplasm and nucleoplasm compared to the nucleolus. Interestingly, the ratio of the calculated mass densities of nucleoplasm and cytoplasm $\rho_{\text{nuc}}/\rho_{\text{cyt}}$=1.003±0.010, is one (within the uncertainty boundaries), which can be explained by the cancellation of the effect of the higher water concentration of nucleoplasm and the presence of the DOPC (which has a lower mass density than proteins) in the cytoplasm.

While the mass density and refractive index values of cytoplasm, nucleoplasm and nucleolus lie in a narrow range of 1.012 g/ml<$\rho$<1.043 g/ml and 1.348<$n$<1.368, respectively, protein aggregates and lipid droplets show – expectedly – striking differences in their mass densities [$\rho_{\text{PA}}^{\text{allW}}$=(1.188±0.023) g/ml, $\rho_{\text{LD}}$=0.90 g/ml, where subscript LD stands for lipid droplets] while exhibiting similarly high refractive index values ($n_{\text{LD}}\approx n_{\text{PA}}^{\text{allW}}$=1.463± 0.015). Naively, this could indicate that the mass density of biological matter for which the measured refractive index values lie within the above stated range, could be roughly estimated as $\rho$=(1.028±0.015) g/ml, ignoring the details about macromolecular composition. And indeed, considering the mass density estimate of zebrafish trunk tissue of [1] [$\rho_{\text{ZF}}$=(1.037±0.006) g/ml], which was based on *in vivo* refractive index measurements of [16] ($n_{\text{ZF}}$=1.3675±0.0034), and the biochemical composition of the tissue, both values are in concordance. This in turn implies that the biological samples with refractive index values higher/lower than the range stated above would need additional information about their composition to avoid under/overestimating the mass density significantly.

**Note S2. Parameter values and cell model when macromolecular composition changes over time:**

In this section, we investigate cases where the macromolecular composition changes over time. For the particular case of a disruption of homeostasis by e.g., a hyper osmotic shock, one may assume the water mass of the cell/tissue to follows a rate equation as $\dot{m}_0(t)=-k_0 m_0(t)+m_0^\infty$, with the solution $m_0(t) = m_0^\infty/k_0 \left[1 - \exp(-k_0 t)\right] + m_0^0 \exp(-k_0 t)$, where $k_0$ is the corresponding rate and the final water mass is given by $m_0(t \to \infty)=m_0^\infty/k_0$. Consequently, the water concentration is given by $c_0(t)=m_0(t)/V_{tot}(t)$, where the total volume of the mixture is $V_{tot}(t)=\sum_i m_i(t)\theta_i + m_0(t)/\rho_0$ and the masses of the individual macromolecules $m_i$ are implicitly given by the rate equations $\dot{m}_i(t) = \sum_{j \neq i} (k_{eff}^{ji} - k_{eff}^{ij}) m_i(t)$, which imply a conserved solute mass. In order to estimate the respective (effective) rates $k_0$ and $k_{eff}^{ij}$, we employ the measurements of [12], where the changes in protein and lipid concentrations of cytoplasm under osmotic shock were determined. We further assume that 1) $k_0 \gg k_{eff}^{ij}$, i.e., the initial change in water concentration is much faster than the macromolecular conversion processes and 2) initially abundant proteins are converted to sugars (glycogen), which subsequently get converted to lipids (DOPC). By setting the initial conditions concordant to the composition of cytoplasm, described earlier, we numerically solve the resulting set of coupled rate equations for the two cases; A) $k_{\text{eff}}^{ij} \neq 0$, i.e., macromolecular conversions, as described above and B) $k_{\text{eff}}^{ij} = 0$, i.e., no macromolecular conversions, using the NDSolve function in Mathematica [17]. The results are shown in Fig. 1, the effective rates and initial conditions are stated in Table S3, and the respective concentrations over time are shown in Fig. S2.

As expected, the resulting correlative mass density-refractive index relationship is non-linear for case A and linear for case B. While the difference between the two cases is not significant for the refractive index range stated above (1.348<$n$<1.368), i.e., at high water concentrations, the effect of the (slow) macromolecular conversion becomes more apparent for lower water concentrations. In fact, the difference of the final refractive indexes is marginal ($n_A$=1.402 ±0.004, $n_B$=1.402 ±0.005), but the difference in the mass densities is immanent [$\rho_A$=(1.107±0.013) g/ml, $\rho_B$=(1.070±0.013) g/ml]. However, while the concentration changes of cytoplasmic proteins and lipids under osmotic shock were chosen to be roughly in line with the measured values, stated in [12], the particular choices of (effective) conversion rates might not reflect physiologically relevant conditions and should be rather seen as an illustrative example.

**Note S3. Derivation of the density estimated from the SBS spectrum:**

From the Clausius–Mossotti relation, the relationship between density and the dielectric constant is:

$$\frac{\varepsilon-1}{\varepsilon+2} = \frac{N\alpha}{3\varepsilon_0} = \frac{\rho\alpha}{3\varepsilon_0}, \quad (S1)$$

where $\varepsilon$ is the dielectric constant of the material, $\varepsilon_0$ is the permittivity of free space, $N$ is the number density of the molecules, $\alpha$ is the molecular polarizability, and $\rho$ is the mass density. The dielectric constant can then be derived as

$$\varepsilon = \frac{1+2\rho\frac{\alpha}{3\varepsilon_0}}{1-\rho\frac{\alpha}{3\varepsilon_0}}. \quad (S2)$$

The electrostrictive constant $\gamma_e$, which is estimated as $\rho\partial\varepsilon/\partial\rho$, can be expressed as

$$\gamma_e = \rho\frac{\partial\varepsilon}{\partial\rho} = \frac{\rho\frac{\alpha}{\varepsilon_0}}{(1-\rho\frac{\alpha}{3\varepsilon_0})^2}. \quad (S3)$$

Because from Eq. (S2), $\varepsilon$-1 and $\varepsilon$+2 can be obtained as

$$\varepsilon - 1 = \frac{\rho\frac{\alpha}{\varepsilon_0}}{1-\rho\frac{\alpha}{3\varepsilon_0}},$$
$$\varepsilon + 2 = \frac{3}{1-\rho\frac{\alpha}{3\varepsilon_0}} \quad (S4)$$

$\gamma_e$ can be simplified as

$$\gamma_e = \frac{(\varepsilon-1)(\varepsilon+2)}{3}. \quad (S5)$$

For non-conducting materials, the refractive index $n$ is approximately equal to the square root of the dielectric constant, so the electrostrictive constant has a relationship with the refractive index as

$$\gamma_e = \frac{(n^2-1)(n^2+2)}{3}. \quad (S6)$$

In SBS measurement, the Brillouin gain $g_B$ of the spectrum is

$$g_B = \frac{4\pi\gamma_e^2 v^3}{c^4 \rho v_B \Gamma_B} = \frac{4\pi\gamma_e^2}{c\lambda^3 \rho v_B \Gamma_B}, \quad (S7)$$

where $c$ is the speed of light, $v$ is the incident optical frequency, $\lambda$ is the incident wavelength, and $v_B$, $\Gamma_B$ represent the Brillouin frequency and the linewidth respectively. Combining Eqs. (S6) and (S7), the mass density can be estimated only depends on the SBS spectral information and the refractive index:

$$\rho_{SBS} = \frac{4\pi}{9c\lambda^3} \frac{(n^2-1)^2(n^2+2)^2}{v_B \Gamma_B g_B}, \quad (S8)$$

**REFERENCES**

(1) Möckel, C.; Beck, T.; Kaliman, S.; Abuhattum, S.; Kim, K.; Kolb, J.; Wehner, D.; Zaburdaev, V.; Guck, J., Estimation of the mass density of biological matter from refractive index measurements. *Biophys. Rep.* **2024,** *4* (2), 100156.

(2) Zhao, H.; Brown, P. H.; Schuck, P., On the distribution of protein refractive index increments. *Biophys. J.* **2011,** *100* (9), 2309-2317.

(3) Parkkila, P.; Elderdfi, M.; Bunker, A.; Viitala, T., Biophysical Characterization of Supported Lipid Bilayers Using Parallel Dual-Wavelength Surface Plasmon Resonance and Quartz Crystal Microbalance Measurements. *Langmuir* **2018,** *34* (27), 8081-8091.

(4) Greenwood, A. I.; Tristram-Nagle, S.; Nagle, J. F., Partial molecular volumes of lipids and cholesterol. *Chem. Phys. Lipids* **2006,** *143* (1), 1-10.

(5) ChemSpider *Triglyceride OPO*; CSID: 21376990; April 29, 2024, 2024.

(6) Kim, S.; Swanson, J. M. J., The Surface and Hydration Properties of Lipid Droplets. *Biophys. J.* **2020,** *119* (10), 1958-1969.

(7) ChemSpider *D-GLYCOGEN*; CSID: 388322; July 09, 2024, 2024.

(8) Geddes, R.; Harvey, J. D.; Wills, P. R., The molecular size and shape of liver glycogen. *Biochem. J.* **1977,** *163* (2), 201-209.

(9) Park, J., The Cell: A molecular approach. *YJBM* **2001,** *74* (5), 361.

(10) Biswas, A.; Muñoz, O.; Kim, K.; Hoege, C.; Lorton, B. M.; Shechter, D.; Guck, J.; Zaburdaev, V.; Reber, S., Conserved nucleocytoplasmic density homeostasis drives cellular organization across eukaryotes. In *bioRxiv*, 2023; p 556409.

(11) Beck, T.; van der Linden, L.-M.; Borcherds, W. M.; Kim, K.; Schlüßler, R.; Müller, P.; Franzmann, T.; Möckel, C.; Goswami, R.; Leaver, M.; Mittag, T.; Alberti, S.; Guck, J., Optical characterization of molecular interaction strength in protein condensates. *Mol. Biol. Cell.* **2024,** *35* (12), ar154.

(12) Liu, X.; Oh, S.; Kirschner, M. W., The uniformity and stability of cellular mass density in mammalian cell culture. *Front. Cell Dev. Biol* **2022,** *10*, 1017499.

(13) Schlüßler, R.; Kim, K.; Nötzel, M.; Taubenberger, A.; Abuhattum, S.; Beck, T.; Müller, P.; Maharana, S.; Cojoc, G.; Girardo, S., Correlative all-optical quantification of mass density and mechanics of subcellular compartments with fluorescence specificity. *Elife* **2022,** *11*, e68490.

(14) Century, T. J.; Fenichel, I. R.; Horowitz, S. B., The concentrations of water, sodium and potassium in the nucleus and cytoplasm of amphibian oocytes. *J. Cell Sci.* **1970,** *7* (1), 5-13.

(15) Päuser, S.; Zschunke, A.; Khuen, A.; Keller, K., Estimation of water content and water mobility in the nucleus and cytoplasm of Xenopus laevis oocytes by NMR microscopy. *Magn. Reson. Imaging* **1995,** *13* (2), 269-276.

(16) Kolb, J.; Tsata, V.; John, N.; Kim, K.; Möckel, C.; Rosso, G.; Kurbel, V.; Parmar, A.; Sharma, G.; Karandasheva, K.; Abuhattum, S.; Lyraki, O.; Beck, T.; Müller, P.; Schlüßler, R.; Frischknecht, R.; Wehner, A.; Krombholz, N.; Steigenberger, B.; Beis, D.; Takeoka, A.;

Blümcke, I.; Möllmert, S.; Singh, K.; Guck, J.; Kobow, K.; Wehner, D., Small leucine-rich proteoglycans inhibit CNS regeneration by modifying the structural and mechanical properties of the lesion environment. *Nat. Commun.* **2023,** *14* (1), 6814.

(17) Inc., W. R., *Mathematica Version 12.2*. Wolfram Research. Inc.: Champaign, 2020.